\documentclass{PoS}

\newcommand{\order}[1]{\mathcal{O}(#1)}
\usepackage[utf8]{inputenc}
\usepackage{chngpage}
\usepackage{amsmath}

\graphicspath{{figs/}}

\title{$B_s\to D_s^{(*)}l\nu$ Form Factors with Heavy HISQ Quarks}

\ShortTitle{$B_s\to D_s^{(*)}l\nu$ Form Factors with Heavy HISQ Quarks}

\author{\speaker{E.~McLean}\\
  SUPA, School of Physics and Astronomy, University of Glasgow, Glasgow, G12 8QQ, UK\\
        E-mail: \email{e.mclean.1@research.gla.ac.uk}}
\author{C.~T.~H. Davies\\
    SUPA, School of Physics and Astronomy, University of Glasgow, Glasgow, G12 8QQ, UK\\}
\author{A.~T.~Lytle\\
  INFN, Sezione di Roma Tor Vergata, Via della Ricerca Scientifica 1, 00133 Roma RM, Italy\\}
\author{J.~Koponen\\
  INFN, Sezione di Roma Tor Vergata, Via della Ricerca Scientifica 1, 00133 Roma RM, Italy\\}
\author{HPQCD Collaboration\\
http://www.physics.gla.ac.uk/HPQCD/ \\}

\abstract{ We present progress on an ongoing calculation of the $B_s\to D_s^{(*)} l \nu$ form factors calculated on the $n_f=2+1+1$ MILC ensembles and using the Highly Improved Staggered Quark action for all valence quarks. We perform the calculation at a range of $b$ quark masses (and lattice spacings) so that we can extrapolate to the physical $b$-quark mass.}

\FullConference{The 36th Annual International Symposium on Lattice Field Theory - LATTICE2018\\
		22-28 July, 2018\\
		Michigan State University, East Lansing, Michigan, USA.}

\begin{document}

\section{Introduction}

Theoretical determinations of $B$ weak decay form factors are important inputs to searches for new physics in the flavour sector. By combining theoretical predictions and experimental branching fractions for decays such as $B\to D^*l\nu$ and $B\to Dl\nu$, one can deduce the CKM element $|V_{cb}|$. Obtaining this to high precision is required to check the unitarity of the CKM matrix, an important test of the standard model (SM). 

There is also interest in $b \to c$ transitions due to a number of anomalies (tension between experiment and the SM) in observables relating to $b\to c$ semileptonic decays. For example there are persistent discrepancies between experimentally observed values and SM predictions for the ratios $R(D^{(*)}) = \mathcal{B}(\bar{B} \rightarrow D^{(*)}\tau^- \bar{\nu}_{\tau})/\mathcal{B}(\bar{B}\to D^{(*)} l^- \bar{\nu}_l)$ ($l=e$ or $\mu$) \cite{PhysRevLett.109.101802}.

Previous calculations have determined $B\to D^{(*)}$ form factors \cite{Harrison:2017fmw,Bailey:2014tva,Lattice:2015rga,Na:2015kha} along with  $B_s\to D_s^{(*)}$ form factors \cite{Harrison:2017fmw,Monahan:2017uby}. $B_{(s)}\to D_{(s)}^*$ results have all been limited to the zero recoil point. An unfortunate feature of each of these results is that each uses a formalism that relies on perturbation theory for the normalization of currents. Hence they contain matching errors of size $\order{\alpha_s^2}$. The NRQCD calculations away from zero recoil also have systematic errors from the truncation of non-relativistic expansions of currents.

In this study we use a pure HISQ approach, where currents can be normalized non-perturbatively. Hence our results do not suffer from matching errors. In our approach we perform the simulation at a number of unphysically light $b$ quark masses (we will simply refer to as {\it{heavy masses}} here), and extrapolate to the physical $b$ mass. By using many heavy masses we can model both form factor dependence on the heavy mass, and the discretization effects associated with $am_h$. It also enables us to obtain lattice data throughout the entire $q^2$ range of the decay, since masses lighter than the $b$ correspond to smaller values for $q^2_{\text{max}}$.

This approach has been shown to work for calculating decay constants \cite{McNeile:2011ng,McNeile:2012qf}, and is currently being used for computing other form factors for the $B_c\to \eta_c$ and $B_c\to J/\psi$ decays \cite{Colquhoun:2016osw}.

In this work, we choose to study only the $B_s\to D_s^{(*)} l\nu$ decays rather than $B\to D^{(*)}l\nu$, since this is a simpler lattice calculation. Light valence quarks are computationally more expensive and contribute more noise to lattice data, studying $B_s\to D_s^{(*)}$ avoids this. The Chiral perturbation theory required to perform extrapolations in light quark mass is also more straightforward in $B_s\to D_s^{(*)}$ compared to $B\to D^{(*)}$ \cite{Laiho:2005ue}.

The $B_s\to D_s^{(*)}$ decays are also phenomenologically interesting in their own right. Experience from previous lattice calculations tells us that the form factors under study are insensitive to the spectator quark mass (see fig. 2 in \cite{Harrison:2017fmw}, fig. 14 in \cite{Monahan:2017uby}). Therefore $B_s\to D_s^{(*)} \simeq  B\to D^{(*)}$ to a good approximation.

Experimental data for $B_s\to D_s^{(*)}$ branching fractions will become available in the future. Then, the theoretical determination given here, along with the experimental data, will supply further SM tests (via comparison of ratios analogous to $R(D^{(*)}$)), and another channel for $|V_{cb}|$ determination.

\section{Calculation Details}

Our two goals are:
\begin{itemize}
    \item
    Deduce the single $B_s\to D_s^{*}$ form factor that contributes at the zero recoil point, $h^s_{A_1}(q^2_{max})$.
  \item
    Deduce the two $B_s\to D_s$ form factors, $f^s_{0}(q^2),f^s_{+}(q^2)$, throughout the physical $q^2$ range, $0 < q^2 < (M_{B_s}-M_{D_s})^2$, where $q^2$ is the momentum transfer.
\end{itemize}
At arbitrary heavy mass (so substituting $m_b$ with $m_h$ and $B_s$ with $H_s$), these form factors are related to current matrix elements via

\begin{align}
      f^s_0(q^2) &= { m_h - m_c \over M_{H_s}^2-M_{D_s}^2 } \langle D_s | S | H_s \rangle,
      \\
      f^s_+(q^2) &= {1\over 2M_{H_s}} { \delta^M  \langle D_s | S | H_s \rangle - q^2 Z_{V} \langle D_s | V_0 | H_s \rangle\over \underline{p}^2_{D_s} }, \\ \nonumber &\quad (\delta^M = (m_h-m_c)(M_{H_s}-M_{D_s})), \\ \nonumber \\
      h^s_{A_1}(q^2_{\text{max}}) &= 2\sqrt{M_{H_s}M_{D_s^*}} Z_{A} \langle D_s^* | A_k | H_s \rangle.
\end{align}

We are here keeping the $m_h$ dependence of the form factors implicit. The currents are all local currents in the HISQ formalism: $S = \bar{\psi}_c\psi_h$, $V_0=\bar{\psi}_c \gamma^0\gamma^5 \psi_h$, $A_k = \bar{\psi}_c \gamma^5 \gamma^k \psi_h$.

We obtain these current matrix elements at varying $a$ and $m_h$ (and $q^2$ for $V_0,S$), by generating correlation functions from lattice simulations on a number of second generation $2+1+1$ MILC ensembles containing HISQ sea quarks \cite{Follana:2006rc,Bazavov:2012xda}. We use the HISQ action for all valence quarks. Parameters for each of the ensembles used are given in table \ref{table:ensembles}. 2- and 3-point correlation functions are computed and fed into multiexponential Bayesian fits, following the methodology of e.g. \cite{Colquhoun:2015mfa}, from which we determine the matrix elements. In the case of $A_k$, we perform the simulation at three spatial polarizations $k = x,y,z$, and corresponding $D_s^*$ polarizations, and take the average.

The local HISQ scalar current $S$ is a partially conserved current, hence is absolutely normalized and requires no normalization constant \cite{Na:2010uf}. The same is not true for $V_0$ and $A_k$. To find their normalizations $Z_{V}$ and $Z_{A}$ non-perturbatively, we demand that certain Ward identities are satisfied:
\begin{align}
(M_{H_s} - M_{D_s}) Z_{V} \langle D_s | V_0 | H_s \rangle &= (m_h - m_c) \langle D_s | S | H_s \rangle, \\
M_{H_s} Z_{A} \langle 0| A_0 | H_s \rangle &= (m_h + m_c) \langle 0 | P | H_s \rangle.
\end{align}
The matrix elements in the first equation are evaluated at zero recoil. The matrix elements $\langle 0 | P | H_s \rangle$ $\langle 0| A_k | H_s \rangle$ were also computed from multiexponential Bayesian fits of lattice data. Here we are leveraging the fact that the local scalar $S$ and pseudoscalar $P=\bar{\psi}_c \gamma^5 \psi_h$ currents are absolutely normalized in HISQ \cite{Donald:2013pea}.

\begin{table}
  \begin{adjustwidth}{-.5in}{-.5in} 
  \begin{center}
    \begin{tabular}{c c c c c c c c c}
      \hline
      handle & $a/$fm  & $N_x^3\times N_t$ & $am_l$ & $am_s$ & $am_c$ & $am_s^{{val}}$ & $am_c^{{val}}$ & $am^{{val}}_h$ \\[0.5ex]
      \hline
      \bf{fine} & 0.0884(6) & $32^3\times96$ & 0.0074 & 0.037 & 0.440 & 0.0376 & 0.45
      & 0.5, 0.65, 0.8 \\ [1ex]
      \bf{superfine} & 0.05922(12) & $48^3\times144$ & 0.0048 & 0.024 & 0.286 & 0.0234 &
      0.274 & 0.427, 0.525, 0.65, 0.8 \\ [1ex]
      \bf{ultrafine} & 0.04406(23) &  $64^3\times192$ & 0.00316 & 0.0158 & 0.188 & 0.0165
      & 0.194 & 0.5, 0.65, 0.8 \\ [1ex]
      \hline
    \end{tabular}
  \caption{Parameters for gluon ensembles \cite{Bazavov:2012xda}. $a$ is the lattice spacing, values found in \cite{Dowdall:2013rya}. $N_x$ is the spatial extent and $N_t$ the temporal extent of the lattice. Light, strange and charm quarks are included in the sea, their masses are given in columns 4-6, and the valence quark masses in columns 7-9. These are tuned in \cite{Chakraborty:2014aca}. We use a number of heavy quark masses to assist the extrapolation to physical $b$ mass.}
  \label{table:ensembles}
  \end{center}
  \end{adjustwidth}
\end{table}

Once we have obtained the form factors at varying $m_h,q^2$ and $a$, we can perform extrapolations to $m_h=m_b$ and (in the $f_{0}^s$, $f_{+}^s$ case) all $q^2$. In the $B_s\to D_s^*$ case we use the fit form
\vspace{-0.03cm}
\begin{align}
  \label{eq:fitfun_hA1}
  h^s_{A_1}(q^2_{max}) =& \, \eta_A \left( 1 - {l_V\over 4(M_{D_s}-\bar{\Lambda})^2} + { l_A \over (M_{H_s}-\bar{\Lambda})(M_{D_s}-\bar{\Lambda})}\, - { l_P \over 4(M_{H_s}-\bar{\Lambda})^2} \right) \\
  \nonumber
  + &\sum_{i,j+k\neq 0}^{2,2,2}
  b_{ijk} \left({\Lambda_{\text{QCD}}\over M_{\eta_h} }\right)^{i} \left({ am_h \over \pi }\right)^{2j} \left({ am_c \over \pi }\right)^{2k} \\
  \nonumber
  + &\left( c_0 + c_1 {\Lambda_{\text{QCD}}\over M_{\eta_h}}
  + c_2 \left( \left({am_c\over \pi}\right)^2 + \left({ am_h \over \pi }\right)^2 \right) \right) \left({ M_{\eta_c} - M_{\eta_c}^{\text{phys}} \over 1\text{GeV}}\right) \\
  \nonumber
  + &\left( s_0 + s_1 {\Lambda_{\text{QCD}}\over M_{\eta_h}}
  + s_2 \left( \left({am_c\over \pi}\right)^2 + \left({ am_h \over \pi }\right)^2 \right) \right) \left({ M_{\eta_s}^2 - M_{\eta_s}^{\text{phys}2} \over 1\text{GeV}^2}\right)
\end{align}
\vspace{-0.03cm}
The first line is inspired by the continuum heavy quark effective theory (HQET) expression for $h_{A_1}$ \cite{Bailey:2014tva}, with the quark masses replaced with $(M_{Q_s} - \bar{\Lambda}) \simeq m_q$. $\bar{\Lambda} = 0.552$GeV is the minimal renormalon subtracted heavy meson binding energy calculated in \cite{Bazavov:2018omf}. $\eta_A = 0.960$ is the 2-loop HQET-QCD matching factor \cite{Neubert:1996wg}. $l_V, l_A, l_P$ are fit parameters.

The rest of the lines are nuisance parameters, accounting for discretization effects and charm and strange mass mistunings, following the approach of \cite{McNeile:2012qf}. $b_{ijk},c_i,s_i$ are fit parameters.

In the $B_s\to D_s$ case, we choose to perform the extrapolation of the ratios $f_{0,+}^s/f_{H_c}\sqrt{M_{H_c}}$, since discretization effects largely cancel in this ratio. We use the following fit form:
\begin{align}
      {f^s_{0,+}(q^2)\over f_{H_c}\sqrt{M_{H_c}}} =& {1\over 1 - q^2/(M_{H_c^{0,*}})^2} \left( 1 + p \log\left({M_{\eta_h}\over M_{\eta_c}}\right) \right) \sum_{l=0}^2 A_l z(q^2)^l \\ \nonumber\\
      \nonumber
      A_l = & \sum_{i,j,k=0}^{2,2,2} b_{ijkl} \left({\Lambda_{\text{QCD}}\over M_{\eta_h} }\right)^{i} \left({ am_h \over \pi }\right)^{2j} \left({ am_c\over \pi }\right)^{2k} \\
      \nonumber
      + &\left( c_{0l} + c_{1l} {\Lambda_{\text{QCD}}\over M_{\eta_h}}
      + c_{2l} \left( \left({am_c\over \pi}\right)^2 + \left({ am_h \over \pi}\right)^2 \right) \right) \left({ M_{\eta_c} - M_{\eta_c}^{\text{phys}} \over 1\text{GeV} }\right) \\
      \nonumber
      + &\left( s_{0l} + s_{1l} {\Lambda_{\text{QCD}}\over M_{\eta_h}}
      + s_{2l} \left( \left({am_c\over \pi}\right)^2 + \left({ am_h \over \pi}\right)^2 \right) \right) \left({ M_{\eta_s}^2 - M_{\eta_s}^{\text{phys}2} \over 1\text{GeV}^2 }\right)
      \\ \nonumber \\
      \nonumber
      z(q^2) &= {\sqrt{t_+ - q^2} - \sqrt{t_+ - t_0} \over \sqrt{t_+ - q^2} + \sqrt{t_+ - t_0}} \quad,\quad t_+ = (M_{H_s}+M_{D_s})^2 \quad,\quad t_0 = (M_{H_s}-M_{D_s})^2
\end{align}
This is a version of the BCL parameterization \cite{Bourrely:2008za}, augmented with nuisance parameters for disretization effects and mass mistunings. The first term accounts for subthreashold poles in the form factors. The second term accounts for any possible logarithmic dependence on the $\eta_h$ mass. $p,b_{ijkl},c_{il},s_{il}$ are fit parameters.

\section{Results}
  
The below results are preliminary. They do not yet contain the full set of statistics we plan to include in the study, so statistical errors will be reduced in the final work. The results do not take into account the systematic error associated with unphysically heavy pions. We will include data at physical pion mass in the future to test for any possible effect. Since there are no valence light quarks, these effects should be small. The results do not take account of errors due to finite volume effects and isospin breaking effects. These are also expected to be small.

\begin{figure}[ht]
  \begin{center}
  \includegraphics[width=0.65\textwidth]{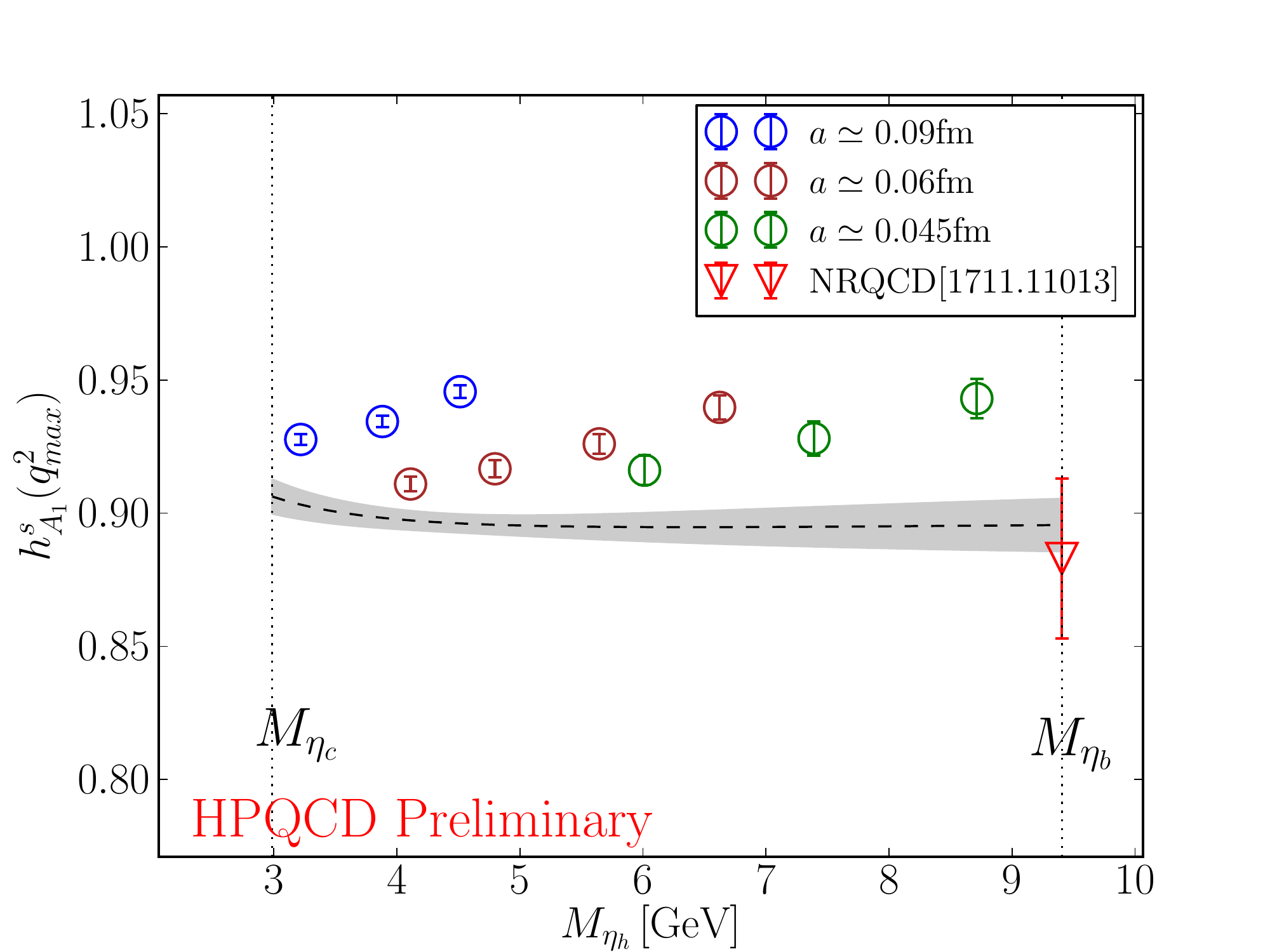}
  \caption{$h_{A_1}^s(q^2_{max})$ against $M_{\eta_h}$ (this can be seen as a proxy for $m_h$). The grey band shows the result of the extrapolation to the continuum. The NRQCD value is from \cite{Harrison:2017fmw}.\label{fig:hA1vsmetah}}
  \end{center}
\end{figure}

The results of the extrapolation of $h^s_{A_1}(q^2_{max})$ is given in fig.\ref{fig:hA1vsmetah}. For comparison we have included the result for $h_{A_1}^s(q^2_{max})$ from an approach using the non-relativistic QCD (NRQCD) action for the $b$ quark \cite{Harrison:2017fmw}. Our result is both in agreement with the NRQCD value, and is considerably more precise.

\begin{figure}[ht]
  \begin{center}
  \includegraphics[width=0.65\textwidth]{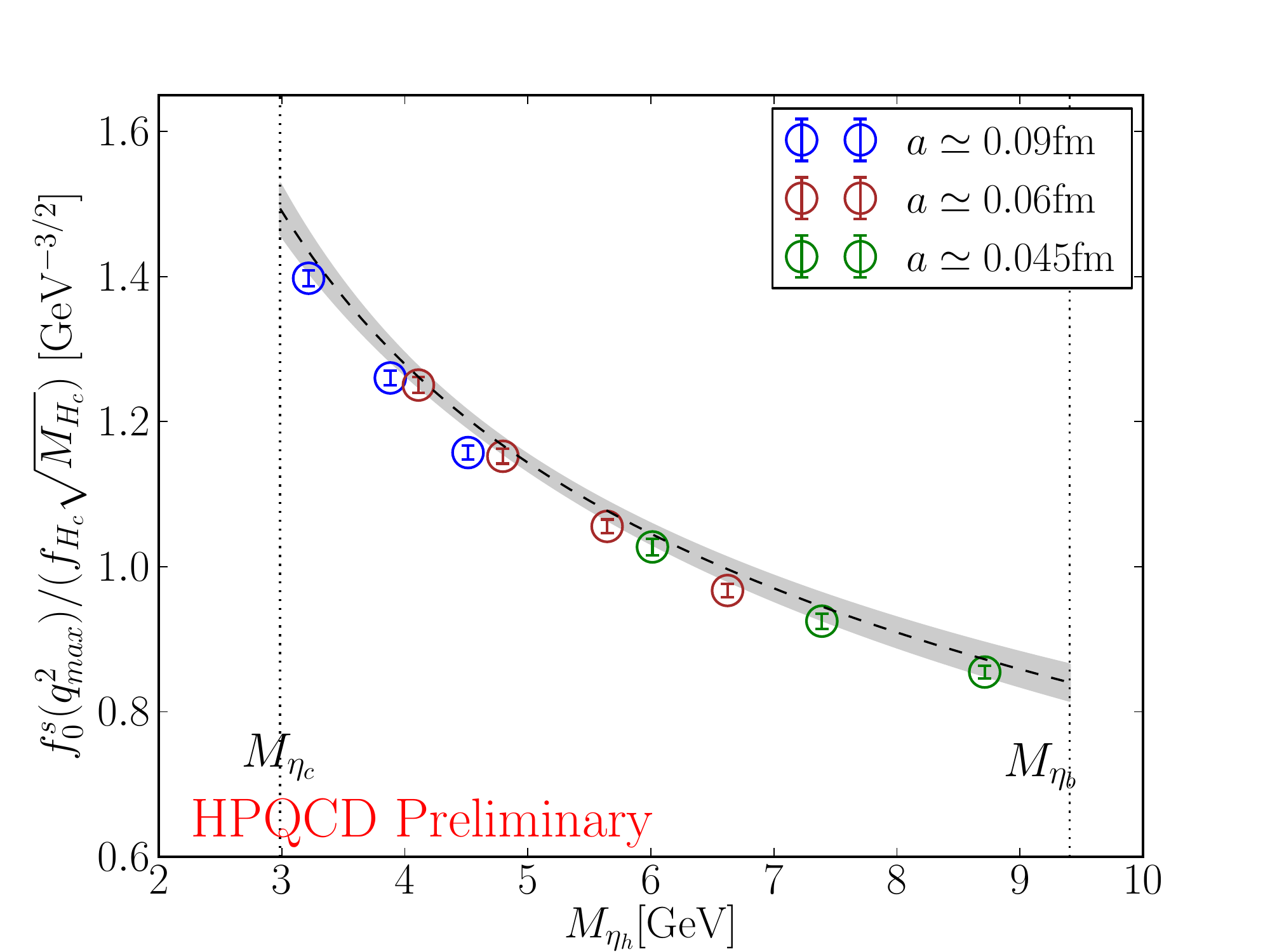}
  \caption{ Extrapolation of $f^s_{0}(q^2_{max})/f_{H_c}\sqrt{M_{H_c}}$ through $M_{\eta_h}$ (a proxy for $m_h$). The grey band shows the result of the extrapolation at continuum.}
  \label{fig:f0fHc_vsmh}
  \end{center}
\end{figure}

\begin{figure}[ht]
  \begin{adjustwidth}{-.5in}{-.5in}
  \begin{center}
  \includegraphics[width=1.00\textwidth]{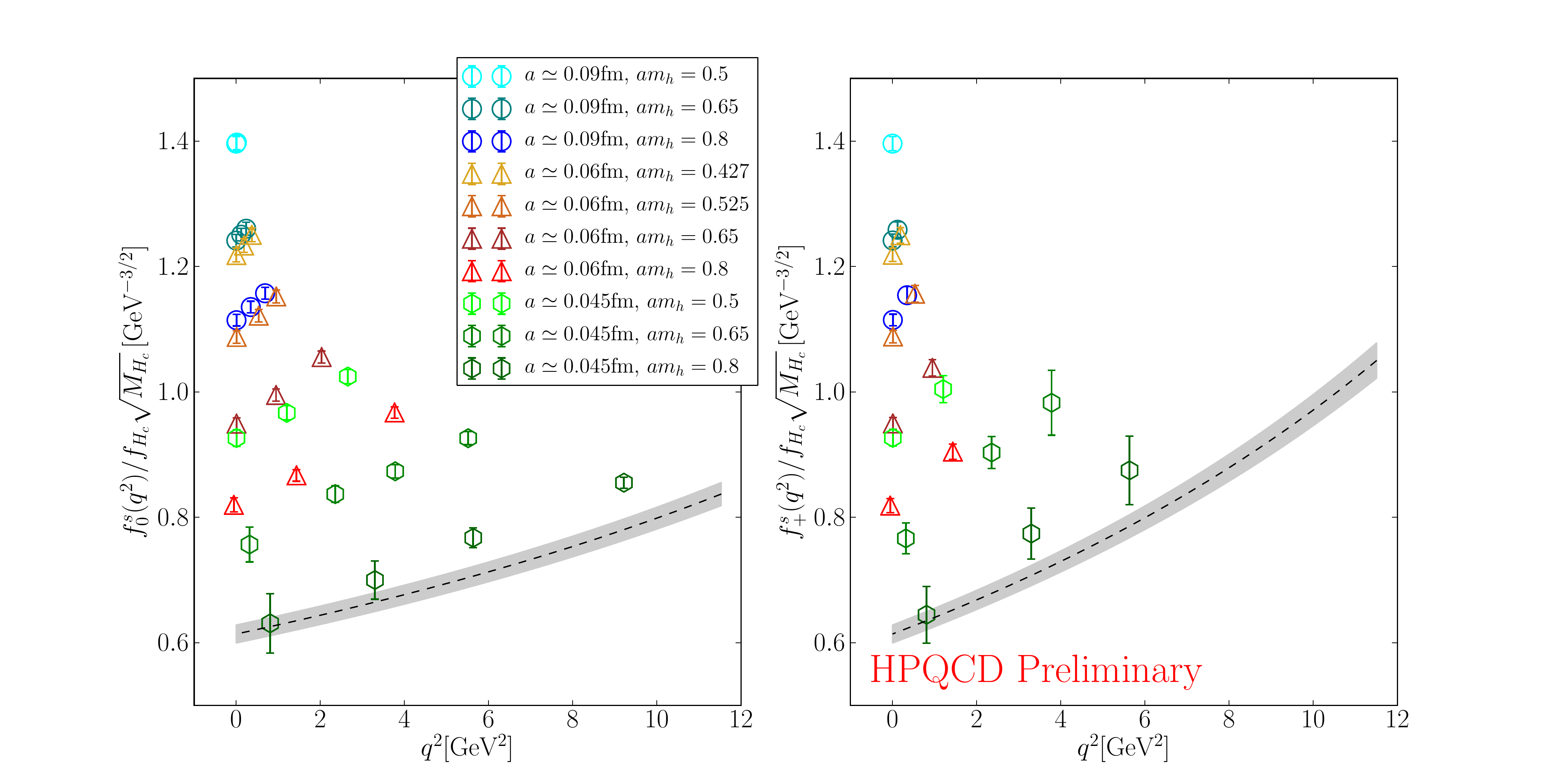}
  \caption{Extrapolation of $f^s_{0,+}(q^2)/f_{H_c}\sqrt{M_{H_c}}$, throughout the $q^2$ range. The grey band shows the extrapolation result in the continuum limit.}
  \label{fig:f0fpdc_vsq2}
  \end{center}
  \end{adjustwidth}
\end{figure}

The results of the $f^s_{0,+}(q^2)/f_{H_c}\sqrt{M_{H_c}}$ extrapolation at $q^2_{\text{max}}$ is given in fig. \ref{fig:f0fHc_vsmh}. The results throughout the entire physical $q^2$ range is given in fig. \ref{fig:f0fpdc_vsq2}.

\section{Conclusion}

We have obtained the $B_s\to D_s^*$ form factor at zero recoil $h^s_{A_1}(q^2_{max})$, and the $B_s\to D_s$ form factors $f_{0,+}^s(q^2)$ throughout the physical $q^2$ range. The method was fully relativistic and contains no perturbative matching errors.. Our results already show improved accuracy over values from NRQCD that must include perturbative matching uncertainties. We plan further improvements to our results in the future, meanwhile they demonstrate the potential for successful application of the heavy HISQ quarks approach for form factors.

{\bf{Acknowledgements}} We are grateful to MILC for the use of their gluon field ensembles. This work was supported by the UK Science and Technology Facilities Council. The calculations used the DiRAC Data Analytic system at the University of Cambridge, operated by the University of Cambridge High Performance Computing Service on behalf of the STFC DiRAC HPC Facility (www.dirac.ac.uk). This is funded by BIS National e-infrastructure and STFC capital grants and STFC DiRAC operations grants.

\bibliographystyle{woc}
\bibliography{skeleton}

\end{document}